\documentclass[12pt,preprint]{aastex}
\usepackage{epsfig}

\shortauthors{Chang \& Cui} \shorttitle{Variable X-ray Absorption Lines in 
Cyg X-1}

\begin{document}

\title{Dramatic Variability of X-ray Absorption Lines in the Black Hole 
Candidate Cygnus X-1}

\author{Chulhoon Chang\altaffilmark{1} and Wei Cui\altaffilmark{1}}
\affil{Department of Physics, Purdue University, West Lafayette,
IN 47907}
\altaffiltext{1}{Email: chang40@physics.purdue.edu, cui@physics.purdue.edu}

\begin{abstract}

We report results from a 30 ks observation of Cygnus X-1 with the High Energy
Transmission Grating Spectrometer (HETGS) on board the {\em Chandra X-ray 
Observatory}. Numerous absorption lines were detected in the HETGS spectrum. 
The lines are associated with highly ionized Ne, Na, Mg, Al, Si, S, and Fe, 
some of which have been seen in earlier HETGS observations. Surprisingly, 
however, we discovered dramatic variability of the lines over the duration 
of the present observation. For instance, the flux of the Ne X line at 
12.14 \AA\ was about $5 \times 10^{-3}$ photons cm$^{-2}$ s$^{-1}$ in the 
early part of the observation but became subsequently undetectable, with a 
99\% upper limit of $0.06 \times 10^{-3}$ photons cm$^{-2}$ s$^{-1}$ on the 
flux of the 
line. This implies that the line weakened by nearly two orders of magnitude 
on a timescale of hours. The overall X-ray flux of the source did also vary 
during the observation but only by 20--30\%. For Cyg X-1, the absorption 
lines are generally attributed to the absorption of X-rays by ionized stellar 
wind in the binary system. Therefore, they may provide valuable diagnostics 
on the physical condition of the wind. We discuss the implications of the
results.

\end{abstract}

\keywords{binaries: general --- black hole physics --- stars: winds,outflows 
--- stars: individual (Cygnus X-1) --- X-rays: binaries}

\section{Introduction}

Cygnus X-1 is the first dynamically-determined black hole system (Webster \& 
Murdin 1972; Bolton 1972). It is in a binary system with a massive O9.7 Iab 
supergiant, and the orbital period was determined optically to be 5.6 days. 
Cyg~X-1 is thus intrinsically different from the majority of known black 
hole candidates (BHCs) whose companion stars are much less massive 
(see review by McClintock \& Remillard 2006). Curiously, those that have a 
high-mass companion (including Cyg~X-1, LMC~X-1 and LMC~X-3) are all 
persistent X-ray sources, while those that have a low-mass companion are 
exclusively transient sources. Perhaps, stellar wind from the companion star 
plays a significant role in this regard (Cui, Chen, \& Zhang 1997).

Unlike transient BHCs, in which mass accretion is mediated by the companion
star overfilling its Roche-lobe, Cyg X-1 is thought to be a wind-fed system.
In this case, however, the wind is thought to be highly focused toward the
black hole, because the companion star is nearly filling its Roche lobe
(Gies \& Bolton 1986). The observed orbital modulation of the X-ray flux
(Wen et al. 1999; Brocksopp et al. 1999a) has provided tentative evidence 
for wind accretion, because it is probably caused by varying amount of 
absorption through the wind (Wen et al. 1999). On the other hand, it is 
generally believed that an accretion disk is also present, based on the
presence of an ultra-soft component, as well as Fe K$_{\alpha}$ emission 
line in the X-ray spectrum (e.g., Ebisawa et al. 1996; Cui et al. 1998).

Cyg X-1 is probably still the most studied BHC. It is a fixture in the target 
list for all major X-ray missions. Much has been learned from modeling the 
X-ray continuum of the source, as well as from studying its X-ray variability.
A recent development is the availability of high-resolution X-ray data, which 
may shed further light on the accretion process and the environment within 
the binary system. Cyg X-1 has been observed on many occasions with the High 
Energy Transmission Grating Spectrometer (HETGS) on board the {\em Chandra 
X-ray Observatory} and the Reflection Grating Spectrometer on board the 
{\em XMM-Newton Observatory}. The high-resolution spectra have revealed the 
presence of numerous absorption lines that are associated with highly ionized 
material (Marshall et al. 2001; Schulz et al. 2002; Feng et al. 2003; Miller 
et al. 2005). 

In this work, we report the detection of absorption lines, some of which have
been seen previously but are much stronger here, and, more surprisingly, the 
discovery of dramatic variability of the lines, based on data from our
HETGS/Chandra observation of Cyg X-1 during its 2001 state transition. The 
fluxes of some of the lines varied by nearly two orders of magnitude over 
the duration of the observation, while the overall flux of the source varied 
only mildly.

\section{Observations and Data Reduction}

Cyg X-1 made a rare transition between the low-hard state and the high-soft 
state, as seen by the All-Sky Monitor (ASM) on the {\em Rossi X-ray Timing 
Explorer} (RXTE), about five years after a similar episode in 1996 (Cui et
al. 1997a and 1997b). Figure~1 shows the ASM light curve that covers the 
entire period. In this case, the flux of the source stayed high for $\sim$400 
days, about twice as long as in 1996. Otherwise, the two episodes are very 
similar, including the flux levels of the two states, prominent X-ray flares 
in both states, and rapid transitions.

Triggered by the ASM alert, the source was observed from 2001 October 28 
16:13:52 to October 29 00:33:52 (UT) with the HETGS on {\em Chandra} 
(ObsID \#3407). The HETGS consists of two gratings: Medium Energy Grating 
(MEG) and High Energy Grating (HEG). After passing the gratings, the photons 
are recorded and read out with the spectroscopic array of the Advanced CCD 
Imaging Spectrometer (ACIS). To avoid photon pile-up in the dispersed events, 
we chose to run the ACIS in continuous-clocking (CC) mode. We also applied 
a spatial window 
around the aim point to accept every $10^{th}$ event in the zeroth order, 
to prevent telemetry saturation yet still have a handle on the position of 
the zeroth-order image for accurate wavelength calibration.

The {\em Chandra} data were reduced and analyzed with the standard {\it CIAO} 
analysis package (version 3.2). Following the {\it CIAO 3.2} Science 
Threads\footnote{See http://asc.harvard.edu/ciao3.2/threads/index.html}, we 
prepared and filtered the data, produced the Level~2 event file from 
the Level~1 data products, constructed the light curves, and made the spectra 
and the corresponding response matrices and auxilliary response files 
(ARFs). We did have to work around a problem related to the use of 
{\em maskfile} for the CC-mode data, when we were making the ARFs. The 
solution is now included in the {\it CIAO 3.3} Science Threads. To verify 
wavelength calibrations, we compared the plus and minus orders and found good 
agreement on the position of prominent lines. We did not subtract background
from either the light curves or the spectra, because it is not obvious how 
to select background events from the CC-mode data. For a bright source like 
Cyg X-1, however, we expect the effects to be entirely negligible.

In coordination with the {\em Chandra} observation, we also observed Cyg X-1
with the {\em Proportional Counter Array} (PCA) and {\em High Energy X-ray
Timing Experiment} (HEXTE) detectors on board {\em RXTE}. The PCA and HEXTE 
covers roughly the energy ranges of 2--60 keV and 15-250 keV, respectively. 
In this work, we made use of the combined broad spectral coverage of the two
detectors to constrain the X-ray continuum more reliably. The {\em RXTE} 
observation was carried out in six short segments, with exposure times of 
1--3 ks. The data were reduced with the standard {\it HEASOFT} package 
(version 5.2), along with the associated calibration files and background 
models. We followed the usual procedures\footnote{See http://heasarc.gsfc.nasa.gov/docs/xte/recipes/cook\_book.html} in preparing, filtering, and reducing 
the data, as well as in deriving light curves and spectra from the 
standard-mode data. While the HEXTE background was directly measured from 
off-source observations, the PCA background was estimated based on the 
background model appropriate for bright sources.

\section{Analysis and Results}

\subsection{Light Curves}

Figure~2 shows the {\em Chandra} light curve of Cyg~X-1 that is made from the 
MEG first-order data (with plus and minus one orders coadded). For comparison,
we have over-plotted the average count 
rate of PCU \#2 for each of the {\em RXTE} exposures. The agreement is quite 
good. Besides the rapid flares and other short-term variability that are 
expected in Cyg X-1, the source also varies significantly on a timescale of 
hours. The MEG count rate rises from a baseline level of roughly 170 counts 
s$^{-1}$ to a peak of roughly 220 counts s$^{-1}$ within 3--4 hours and 
quickly goes back down to the baseline level. We somewhat arbitrarily divided 
the light curve into two time periods (which are labeled as Period~I and 
Period II in Fig. 2) for subsequent analyses. 

\subsection{High-Resolution Spectroscopy}

We analyzed and modeled the HETGS spectra in {\it ISIS} (version 1.2.9) 
(Houck \& Denicola 2000)\footnote{See also http://space.mit.edu/CXC/ISIS/}. 
For this work, we focused only on the first-order spectra, because of the 
relatively poor signal-to-noise ratio of higher-order spectra. For both 
the MEG and HEG, we first co-added the plus and minus orders to produce 
an overall first-order spectrum, again following the appropriate 
{\it CIAO 3.2} science threads. The resolution of the raw data is about 
0.01 {\AA} and 0.005 {\AA} for the MEG and HEG, respectively, which 
represents a factor of 2 over-sampling of the instrumental resolution. We
applied no further binning of the data. Each spectrum was broken into 
3-{\AA} segments for subsequent analyses. Each segment was fitted locally 
with a model that consists of a multi-color disk 
component and a power law for the continuum and negative or positive 
Gaussian functions for narrow absorption or emission features, with the
interstellar absorption taken into account. Such a continuum model is 
typical of BHCs. However, the best-fit continuum differs among the segments 
or between the MEG and HEG, presumably due to remaining calibration 
uncertainties. Since we are only interested in using the {\em Chandra} data 
to study lines, we think that the adopted procedure is justified. We use 
the {\em RXTE} data to more reliably constrain the continuum.

One thing that one notices right away is the presence of many absorption 
lines in the high-resolution spectra. For this work, we consider a feature 
real if it is present both in the MEG and HEG data, with a significance of 
above 4$\sigma$. Figure 3 shows a portion of the MEG first-order spectrum 
for Period I, to highlight the lines detected. No absorption lines were
found at wavelengths above 15 \AA. We should point out 
that we also see the emission-like features at 6.74 {\AA} and 7.96 {\AA},  
which most likely instrumental artifacts associated with calibration 
uncertainties around the Si K and Al K edges (Miller et al. 2005 and 
references therein). We have identified the absorption lines with highly 
ionized species of Ne, Na, Mg, Al, Si, S, and Fe, based mostly on the atomic 
data in Verner et al. (1996) and Behar et al. (2002), but also in 
ATOMDB\footnote{See http://cxc.harvard.edu/atomdb}~1.3.1 for additional 
transitions. 

The line identification 
process involves three steps: (1) a transition is considered a candidate if
the theoretical wavelength is within 0.03 {\AA} of the measured value; 
(2) in cases where multiple candidates exist based on (1), the most 
probably one (based both on the oscillator strength of the transition and 
the relative abundance of the element) is chosen; and (3) consistency
check is made to avoid the identification of a line when more probable 
transitions of the same ion are not seen. The last step is critical. For
instance, we initially associated the lines at 10.051 {\AA} and 11.029 {\AA}
with Fe XVIII $2s^2 2p^5$--$2s^2 2p^4 6d$ and Fe XVII 
$2s^2 2p^6$--$2s 2p^6 4p$, respectively, because, assuming solar 
abundances, they are expected to be stronger than Na XI $1s$--$2p$ and
Na X $1s^2$--$1s 1p$, which were also viable candidates from Step~1. However,
we did not detect other more probable transitions associated with Fe XVIII 
and Fe XVII, which made the identifications highly unlikely. We had to 
conclude that Fe is under-abundant by a factor of 2--3, at minimum, so 
that we could associate the lines with Na instead. If so, the line
at 7.480 {\AA} would more likely be associated with Mg XI $1s^2$--$1s 4p$, 
as opposed to Fe XXIII $2s^2$--$2s 5p$, which we initially identified. The
problem with this is that we saw no hint of Mg XI $1s^2$--$1s 3p$, which 
is more probable.
Therefore, we had to conclude similarly that Mg is also under-abundant by
at least a similar amount (but not by too much, as constrained by other Mg
lines).

Table~1 show all of the absorption lines that we detected and identified. 
The flux and equivalent width (EW) of each line shown were derived from 
the best-fit Gaussian for the line, as well as the local continuum around 
the line for the latter. Note that in a few cases our identifications are 
different from those in the literature. They include: Fe~XXII at 8.718 {\AA} 
and Fe~XXI at 9.476 {\AA} (cf. Marshall et al. 2001; Miller et al. 2005); 
Fe~XX at 10.12 {\AA}, Na X at 11.0027 {\AA}, and Fe~XX at 12.82 {\AA} 
(cf. Marshall et al. 2001); and  Fe~XXI 11.975 {\AA} (cf. Miller et al. 2005).
Only about half of the absorption lines that we detected have been seen
previously in Cyg X-1 (Marshall et al. 2001; Schulz et al. 2002; Feng et 
al. 2003; Miller et al. 2005). Many of these lines are much stronger in 
our case. On the other hand, we detected nearly all of the reported absorption 
lines. The exceptions include: S~XVI at 4.72 \AA\  
(Marshall et al. 2001; Feng et al. 2003), which is detectable only at the 
$3 \sigma$ level in our case (and is thus not included in the table), 
Fe~XIX at 14.53 \AA\ (Miller et al. 2005), which is actually detectable at 
about the $4\sigma$ level here (but just misses our threshold), and Fe~XIX 
at 14.97 \AA\ and Fe~XVIII at 16.01 \AA\ (Miller et al. 2005), whose presence 
is not apparent in our data ($< 3\sigma$). Miller et al. (2005) also 
reported a line at 7.85 \AA\ (Mg~XI) but only at the $3\sigma$ level. The 
line can also be seen in our data at the $4\sigma$ level (but also just 
misses our threshold). Therefore, we already see some indication that 
the absorption lines in Cyg X-1 may be variable from a comparison of 
our results with those published.

Still, it is surprising that almost all of the absorption lines become 
undetectable 
in Period~II. Figure~4 shows the MEG first-order spectrum for this time 
period, which can be directly compared to results in Fig.~3. The only 
exception is the line at 14.608 \AA, which is detected with a significance 
of $5\sigma$ in Period~II. We derived 99\% upper limits on the flux and EW 
of each line seen in Period I but not in Period II. The results are 
also summarized in Table~1, for direct comparison. As an example, we examine
the Ne X line at 12.1339 {\AA} in the two periods. 
The integrated flux of the line is about $5 \times 10^{-3}$ photons cm$^2$ 
s$^{-1}$ in Period I, while its 99\% upper limit for Period II is only 
$0.06 \times 10^{-3}$ photons cm$^2$ s$^{-1}$. Therefore, the line weakened 
by nearly two orders of magnitude in flux over a timescale of merely several 
hours. This is the first time that such dramatic variability of the lines 
has been observed in any BHC. While this is the most extreme case, other
lines also show large variability (see Table~1).

To quantify the column density of each ion required to account for the 
corresponding absorption line detected and its variability, we carried out 
curve-of-growth analysis, following Kotani et al. (2000). The atomic data 
used in the analysis were again taken from Verner et al. (1996), Behar et 
al. (2002), and also ATOMDB 1.3.1 in some cases. The analysis assumes that 
the width of the lines is due entirely to thermal Doppler broadening. For 
resolved lines, we derived the characteristic temperature from the measured 
widths. For unresolved lines, on the other hand, we adopted a temperature 
that would lead to a line width equal to the resolution of the MEG 
(0.023 {\AA} at FWHM). In these cases, therefore, the derived column density 
only represents a {\em lower} limit. The results are shown in Table~1. This
explains why, e.g., the density of Ne~X derived from the 12.144-{\AA} line 
(which is unresolved) is significantly lower than that from the 10.245-{\AA} 
line. Note, however, that the latter is much lower that that from the 
9.727-{\AA} line. We think that the inconsistency arises from the fact that 
the 9.727-{\AA} line is likely a mixture of the Ne line and the Fe~XIX line
at 9.700 {\AA}. Similar inconsistency is also apparent in a few other cases 
(see Table~1), which may originate similarly in line blending. It is also 
worth noting that most lines that we have analyzed fall on the linear part 
of the curve of growth. All {\it resolved} lines at wavelengths 
$\lambda > 11.7$ {\AA} are saturated; so is the unresolved line at 
14.203 {\AA}. To show the degree of variability, we also derived 
a 99\% upper limit on the column density of each of the ions for Period II 
(assuming the same characteristic temperatures).

\subsection{X-ray Continuum}

We now use the {\em RXTE} data to constrain the X-ray continuum of Cyg X-1.
Both the PCA and HEXTE data were used. For the PCA, we used only data from
the first xenon layer of each PCU, which is most accurately calibrated. 
Consequently, the PCA spectral coverage is limited to roughly 2.5--30 keV. 
We relied on the HEXTE data to extend spectral coverage to higher energies. 
The PCA consists of five detector units, known as PCUs. Not all PCUs were 
always operating. For simplicity, we used only data from PCU \#0 and PCU \#2, 
which stayed on throughout the observation, in the subsequent modeling. 
We chose to derive a spectrum for each PCU separately, as well as for each
of the two HEXTE clusters. Residual calibration errors were accounted
for by adding 1\% systematic uncertainty to the data. We also rebinned the 
HEXTE spectra to a signal-to-noise ratio of at least 3 in each bin. The 
individual spectra were then jointly fitted with the same model that 
includes a multi-color disk component, a broken 
power-law that rolls over exponentially beyond a characteristic energy, and 
a Gaussian, taking into account the interstellar absorption. We also multiply 
the model by a normalization factor that is fixed at unity for PCU \#2 but is 
allowed to vary for other detectors, in order to account for any uncalibrated 
difference in the overall throughput among the detectors. The model fits the 
data well for all six segments in the sense that the reduced $\chi^2$ is near
unity (with 169 degrees of freedom).

The spectral shape of Cyg X-1 varied little from one observation to the other.
The best-fit photon indices are $\sim$2.1 and $\sim$1.7 below and above 
$\sim$10 keV. The roll-over energy stays at 20--21 keV and the e-folding
energy roughly at 120--130 keV. Neither the disk component nor the absorption 
column density is well constrained by the data, due to the lack of sensitivity 
(and, to some extent, large calibration uncertainty) at low energies. The 
results can be compared 
directly with those of Cui et al. (1997a), who applied the same empirical 
model to the {\em RXTE} spectra of Cyg X-1 during the 1996 transition. It is 
quite apparent that the spectra here are significantly harder, implying that 
the source was certainly not yet in the true high-soft state (see Cui et al. 
1997a and 1997b for discussions on the ``settling period''). From the 
long-term ASM monitoring data, we can see that Cyg X-1 was brighter during 
our observation than during any of the earlier {\em Chandra} observations 
(Marshall et al. 2001; Schulz et al. 2002; Miller et al. 2005), but not as 
bright as during a later observation (Feng et al. 2003), when the high-soft 
state appears to have been reached. 

\subsection{Photoionization Modeling}

To shed light on the physical properties of the absorber, we carried out a 
photo-ionization calculation with XSTAR version 2.1 kn3\footnote{See http://heasarc.gsfc.nasa.gov/docs/software/xstar/xstar.html}. The underlying assumption
is that the absorber is photo-ionized by the X-ray radiation from the vicinity
of the black hole. The input parameters include the 0.5-10 keV luminosity 
($L_x=3.11 \times 10^{37}$ erg s$^{-1}$, for a distance of 2.5 kpc), 
power-law photon index (2.1), both of which are based on results from 
modeling one of the {\em RXTE} spectra with an assumed $N_H$ value of 
$5.5 \times 10^{21}$ cm$^{-2}$ (Ebisawa et al. 1996; Cui et al. 1998). We 
should note that it is in general risky to extrapolating the assumed 
power-law spectrum to lower energies, because it could severely over-estimate 
the flux there. For the lines of interest here, however, only ionizing photons
with energies $>$ 1 keV contribute and the spectrum of those photons is 
described fairly well by a power law, because the effective temperature of 
the disk component is expected to very low (e.g., Ebisawa et al. 1996; 
Cui et al. 1998).

One of the outputs of the calculation is abundances of the ions of interest, 
as a function of the ionization parameter, $\xi=L_x/nr^2$, where $n$ is the 
density of the absorber and $r$ is the distance of the absorber to the 
source of ionizing photons. Using these abundance curves, we could, in 
principle, constrain $\xi$ to a range that is consistent with the ratio of 
the densities of any two ions of an element. The challenge in practice is,  
as already mentioned, that many of the lines are likely a blend of multiple 
transitions (of comparable probabilities), which makes it difficult to 
reliably determine the densities. Nevertheless, we made an attempt at 
deriving such constraints with the resolved, non-mixed lines. Figure~5 
summarizes the results. The intervals do not all overlap, which implies 
that no single value of $\xi$ could account for all the data. This is
supported by the fact that we have detected all the lines that are expected 
for ionization parameters in the range of roughly $10^{2.5}$--$10^{4.5}$.
If one assumes that the ``absorbers'' are thin shells along the line of 
sight and that they all have the same density, e.g., $n=10^{11}$ $cm^{-3}$ 
(see, e.g., Wen et al. 1999), one would have $10^{11}$ $cm$ $\lesssim r 
\lesssim $ $10^{12}$ $cm$. Compared with the estimated the distance between 
the compact object and the companion star ($\sim 1.4\times 10^{12}$ $cm$; 
LaSala et al. 1998), this would put the ``absorbers'' within the binary 
system. One should, however, take the results with caution, because of, e.g.,
gross over-simplification regarding the geometry of the ``absorbers''.

\section{Discussion}

The observed dramatic variability of the absorption lines might be caused
by a change in the degree of ionization in the wind. Since the overall X-ray
luminosity varied only mildly, we speculate that it probably arises from a 
sudden change in the density of the wind. There is evidence that such a change 
could occur during a state transition or during flares (Gies et al. 2003).
If the moderate decrease in the ionizing flux is accompanied by a more 
dramatic reduction in the density of the wind from Period I to Period II, 
the ionization parameter might increase sufficiently to cause a total 
ionization of the wind in Period II and thus the disappearance (or significant
weakening) of the lines. It is worth noting that in Period I lighter elements
seen are all H- or He-like but Fe is in an intermediate ionization
state (as indicated by the absence of H- or He-like ions), suggesting a high
but not extreme degree of ionization in the period. 
Conversely, a dramatic
increase in the wind density could achieve the same effect. Numerical 
simulations of similar wind-accretion systems (e.g., Blondin, Stevens, 
\& Kallman 1991) have revealed not only a significant jump in the column 
density at late orbital phases ($\gtrsim$ 0.6) that is associated with 
tidal streams but also large variability of the absorbing column. It is
conceivable that Period~II might coincide with a sudden increase in the
column density. Since we found no apparent absorption lines that correspond 
to a lower degree of ionization in Period~II, however, such lines must be 
outside the spectral range covered with our data, in order for the scenario 
to be viable. A quantitative assessment of these scenarios is beyond the
scope of this work.

Many of the absorption lines detected by Miller et al. (2005) are much 
stronger during Period I of our observation (see Table~1). Using only the
lines that are detected with a significance above $5\sigma$, we looked for 
a systematic red- or blue- shift of the lines, following up on the reported 
redshift of the lines by Marshall et al. (2001) based on data taken in the 
low-hard state. The results are summarized in Figure~6 (in the left panel).
In this case, although the lines are still systematically redshifted on 
average, there is not an obvious single-velocity solution. Interestingly, 
if we limit the results only to those lines that were used by Marshall et 
al. (2001), as shown in the right panel of Figure~6, we would arrive at an 
average velocity that is very close to what Marshall et al. reported, 
although the scatter of data points is much larger in our case. On the 
other hand, our observation spans binary orbital phases from 0.85 to 0.92, 
according to the most updated ephemeris (Brocksopp et al. 1999b), while 
Marshall et al.'s covers a phase range of 0.83--0.86. If the redshift of 
the lines is related to the focused-wind scenario advocated by Miller et 
al. (2005), we ought to see a larger (by about 30\%) redshift. Given the 
large uncertainties, as well as the possibility that the wind geometry
might be different for different states, it is difficult to draw any 
definitive conclusions.

Feng et al. (2003) reported the detection of a number of absorption lines 
of asymmetric 
profile, when Cyg X-1 was in the high-soft state, which they interpreted as
evidence for inflows. The same lines are also present in our data during
Period~I and are, in fact, much stronger (except for S XVI, as noted in 
\S~3.2). Figure~7 shows an expanded view of the Si and Mg lines, which are 
the strongest in the group. As is apparent from the figure, the line profile 
can be fitted fairly well by a Gaussian function in all cases. Therefore, 
the lines show no apparent asymmetry here. This also seems to be the case 
for the S and Fe lines, although the statistics of the data are not as good. 
Taken together, our results and Feng et al.'s imply that that the phenomenon 
is either unique to the high-soft state (in which Feng et al. made the 
observation) or is intermittent in nature. We should also note that Feng et 
al's observation was carried out around the superior conjunction (i.e., zero 
orbital phase), where absorption due to the wind is expected to be the 
strongest (e.g., Wen et al. 1999). It is not clear, however, how such 
additonal absorption would lead to an asymmetry in the profile of lines.

No emission lines are apparent in our data. Evidence for weak emission lines 
has been presented (Schulz et al. 2002; Miller et al. 2005) but the 
significance is marginal in all cases. On the other hand, several absorption 
edges are easily detected in our data (see Figs.~3 and 4), as first reported 
and studied in detail by Schulz et al. (2002). The edges can almost certainly 
be attributed to the interstellar absorption.

\acknowledgments{We thank Harvey Tananbaum for approving this DDT observation, 
Norbert Schulz and Herman Marshall for helpful discussion on the pros and cons
of various observing configurations, John Houck for help with the use of 
{\it ISIS} and Tim Kallman for help on using {\it XSTAR}, and David 
Huenemoerder for looking 
into issues related to the {\em Chandra} data products. We acknowledge the 
use of the curve-of-growth analysis script that Taro Kotani has made publicly 
available. We also thank the anonymous referee for a number of useful 
comments that led to significant improvement of the manuscript. 
Support for this work was provided in part by NASA through the
Chandra Award DD1-2011X issued by the Chandra X-ray Observatory Center, which 
is operated by the Smithsonian Astrophysical Observatory for and on behalf of 
NASA under contract NAS8-03060, and through the LTSA grant NAG5-9998. }

\begin{table}
\caption{Detected Absorption Lines}
\tiny
\setlength{\tabcolsep}{1mm}
\begin{tabular}{lllccccccc}\hline\hline
 &Theoretical&Measured&Shift&Flux(I)&Flux(II)&EW(I)&EW(II)&$N_z$(I)&$N_z$(II)\\
Ion and Transition&({\AA})&({\AA})&(km s$^{-1}$)&\multicolumn{2}{c}{($10^{-3}$ph cm$^{-2}$s$^{-1}$)}&({m\AA})&({m\AA})&($10^{-16}$cm$^{-2}$)&($10^{-16}$cm$^{-2}$)\\\hline
S XV 1s$^2$-1s2p&5.039$^b$&5.041(3)&120$\pm$180&1.8(3)&$<$0.9&3.8(7)&$<$1.5&2.5(5)&$\leq$1.0\\
Si XIV 1s-2p&6.1822$^a$&6.189(1)&330$\pm$50&3.0(2)&$<$0.6&6.9(4)&$<$1.0&5.4(3)&$\leq$0.7\\
Si XIII 1s$^2$-1s2p&6.648$^b$&6.657(2)&410$\pm$90&2.6(2)&$<$1.7&6.1(5)&$<$2.9&2.4(2)&$\leq$1.1 \\
Mg XII 1s-3p&7.1062$^a$&7.119(3)&540$\pm$130&1.2(2)&$<$0.9&2.6(5)&$<$1.4&8(2)&$\leq$4.0\\
Al XIII 1s-2p&7.1727$^a$&$7.191(^{+3}_{-2})$&760$^{+130}_{-80}$&1.3(2)&$<$1.7&3.0(5)&$<$2.7&1.6(3)$^d$&$\leq$1.5 \\
Fe XXIII 1s$^2$2s2p-1s$^2$2s6d&7.2646$^c$&7.268(5)&140$\pm$210&1.4(3)&$<$2.0&3.1(6)&$<$3.3&29(6)&$\leq$31\\
Fe XXIII 1s$^2$2s$^2$-1s$^2$2s5p&7.4722$^a$&7.480$(^{+5}_{-4})$&310$^{+200}_{-160}$&0.9(2)&$<$1.4&1.9(5)&$<$2.3&5(1)&$\leq$6.5\\
Fe XXIV 1s$^2$2s-1s$^2$4p&7.9893$^a$&8.004(5)&550$\pm$190&2.2(3)&$<$1.3&4.7(7)&$<$2.2&9(1)&$\leq$4.0\\
Fe XXII 1s$^2$2s$^2$2p-1s$^2$2s$^2$5d&8.0904$^c$&8.096(3)&210$\pm$110&1.0(2)&$<$0.6&2.1(5)&$<$1.0&8(2)$^d$&$\leq$3.6\\
Fe XXII 1s$^2$2s$^2$2p-1s$^2$2s$^2$5d&8.1684$^c$&8.166(3)&-90$\pm$110&1.0(2)&$<$1.1&2.3(5)&$<$1.9&9(2)$^d$&$\leq$7.2\\
Fe XXIII 1s$^2$2s$^2$-1s$^2$2s4p&8.3029$^a$&8.319(2)&580$\pm$70&3.0(3)&$<$0.6&6.6(6)&$<$1.0&6.4(6)&$\leq$0.9\\
Mg XII 1s-2p&8.4210$^a$&8.428(1)&250$\pm$40&5.0(3)&$<$0.4&10.8(6)&$<$0.7&4.7(3)&$\leq$0.3\\
Fe XXI 1s$^2$2s$^2$2p$^2$-1s$^2$2s$^2$2p5d&8.573$^a$&8.577(5)&140$\pm$170&1.3(3)&$<$0.7&2.8(7)&$<$1.2&6(2)&$\leq$2.6\\
Fe XXII 1s$^2$2s$^2$2p-1s$^2$2s2p$_{1/2}$4p$_{3/2}$&8.718$^c$&8.735$(^{+2}_{-3})$&580$^{+70}_{-100}$&2.2(3)&$<$0.5&4.7$(^{+7}_{-6})$&$<$0.9&7(1)&$\leq$1.3 \\
Fe XXI 1s$^2$2s$^2$2p$^2$-1s$^2$2s2p$_{1/2}$2p$_{3/2}$4p$_{3/2}$&8.8254$^c$&8.823$(^{+3}_{-4})$&-80$^{+100}_{-140}$&1.4(3)&$<$0.9&3.0$(^{+7}_{-6})$&$<$1.4&21$(^{+5}_{-4})$&$\leq$9.4 \\
Fe XXII 1s$^2$2s$^2$2p-1s$^2$2s$^2$4d&8.98$^a$&8.978$(^{+1}_{-2})$&-70$^{+30}_{-70}$&1.8(3)&$<$0.9&3.9(6)&$<$1.6&4.6(8)$^d$&$\leq$1.8 \\
Fe XXII 1s$^2$2s$^2$2p-1s$^2$2s$^2$4d&9.07$^a$&9.083$(^{+2}_{-3})$&430$^{+70}_{-100}$&1.9$(^{+3}_{-4})$&$<$0.6&4.0$(^{+6}_{-8})$&$<$1.0&5.2$^{+0.8}_{-1.1}$&$\leq$1.2 \\
Mg XI 1s$^2$-1s2p&9.170$^b$&9.192$(^{+2}_{-1})$&720$^{+70}_{-30}$&6.2(4)&$-$&13.5(9)&$-$&2.9(2)&$-$ \\
Fe XXI 1s$^2$2s$^2$2p$^2$-1s$^2$2s$^2$2p4d&9.356$^a$&9.378(5)&700$\pm$160&1.9(4)&$<$2.8&4.3$^{+1.0}_{-0.9}$&$<$4.8&9(2)&$\leq$10 \\
Fe XXI 1s$^2$2s$^2$2p$^2$-1s$^2$2s$^2$2p4d&9.476$^a$&9.478(1)&60$\pm$30&3.8(3)&$<$0.8&8.6$(^{+8}_{-7})$&$<$1.5&6.6$(^{+7}_{-6})$$^d$&$\leq$1.0 \\
Fe XIX 1s$^2$2s$^2$2p$^4$-1s$^2$2s$^2$2p$^3$($^2$D)5d&9.68$^a$&9.700(4)&620$\pm$120&4.4$(^{+5}_{-6})$&$<$1.8&9(1)&$<$3.1&30(3)&$\leq$9.8\\
Ne X 1s-4p&9.7082$^a$&9.727$(^{+2}_{-3})$&580$^{+60}_{-90}$&2.4$(^{+3}_{-4})$&$<$0.7&5.3$(^{+8}_{-9})$&$<$1.2&24(4)&$\leq$5.0 \\
Fe XX 1s$^2$2s$^2$2p$^3$-1s$^2$2s$^2$2p$^2$($^3$P)4d&9.991$^a$&10.000(1)&270$\pm$30&3.7(4)&$<$0.7&8.3(8)&$<$1.3&5.6(6)$^d$&$\leq$0.8 \\
Na XI 1s-2p&10.0250$^a$&10.051(2)&780$\pm$60&5.6$(^{+5}_{-6})$&$<$2.0&13(1)&$<$3.5&4.0(3)&$\leq$1.0 \\
Fe XX 1s$^2$2s$^2$2p$^3$-1s$^2$2s$^2$2p$^2$($^3$P)4d&10.12$^a$&10.127(3)&210$\pm$90&1.7(4)&$<$0.3&3.8(9)&$<$0.6&2.3(6)$^d$&$\leq$0.4 \\
Ne X 1s-3p&10.2389$^a$&10.245$(^{+3}_{-2})$&180$^{+90}_{-60}$&2.7$(^{+4}_{-5})$&$<$0.5&6(1)&$<$0.9&9(2)&$\leq$1.2 \\
Fe XXIV 1s$^2$2s-1s$^2$3p&10.619$^a$&10.631$(^{+2}_{-1})$&340$^{+60}_{-30}$&9.1(6)&$<$4.6&22(2)&$<$8.9&10(1)&$\leq$3.7 \\
Fe XXIV 1s$^2$2s-1s$^2$3p&10.663$^a$&10.674(3)&310$\pm$80&5.6(6)&$<$3.2&14(2)&$<$6.2&12(2)&$\leq$5.0 \\
Fe XIX 1s$^2$2s$^2$2p$^4$-1s$^2$2s$^2$2p$^3$($^4$S)4d&10.816$^c$&10.818(5)&60$\pm$140&7.0(9)&$<$3.0&18(2)&$<$6.0&14(2)&$\leq$4.3 \\
Fe XXIII 1s$^2$2s$^2$-1s$^2$2s3p&10.981$^a$&10.990(1)&230$\pm$30&5.5(5)&$<$2.7&14(1)&$<$5.6&2.4(2)$^d$&$\leq$0.8 \\
Na X 1s$^2$-1s2p&11.0027$^a$&11.029(2)&720$\pm$50&6.3$(^{+6}_{-7})$&$<$4.4&17(2)&$<$9.2&2.7(4)&$\leq$1.3 \\
Fe XVIII 1s$^2$2s$^2$2p$^5$-1s$^2$2s$^2$2p$^4$($^1$D)4d&11.326$^c$&11.33(1)&100$\pm$260&8(1)&$<$5.1&23$(^{+4}_{-3})$&$<$11&24$(^{+4}_{-3})$&$\leq$11 \\
Fe XXII 1s$^2$2s$^2$2p-1s$^2$2s2p($^3$P$^0$)3p&11.44$^a$&11.431(1)&-240$\pm$30&7.3(6)&$<$2.5&21(2)&$<$5.7&8(1)$^d$&$\leq$1.6\\
Fe XXII 1s$^2$2s$^2$2p-1s$^2$2s2p($^3$P$^0$)3p&11.51$^a$&11.500(3)&-260$\pm$80&4.3$(^{+8}_{-7})$&$<$0.6&12(2)&$<$1.4&22($^{+5}_{-4}$)&$\leq$2.3\\
Fe XXII 1s$^2$2s$^2$2p-1s$^2$2s$^2$3d&11.77$^a$&11.781$(^{+3}_{-1})$&280$^{+80}_{-30}$&12.6(9)&$<$3.8&39(3)&$<$9.3&7.5$^{+1.0}_{-0.9}$&$\leq$1.2\\
Fe XXI 1s$^2$2s$^2$2p$^2$-1s$^2$2s2p$^2$3p&11.975$^c$&11.982(2)&180$\pm$50&9.1(9)&$-$&29(3)&$-$&17($^{+3}_{-2})$&$-$ \\
Ne X 1s-2p&12.1339$^a$&12.144($^{+2}_{-1}$)&250$^{+50}_{-20}$&4.7(7)&$<$0.06&16(2)&$<$0.2&3.9$(^{+7}_{-6})$$^d$&$\leq$0.04 \\
Fe XXI 1s$^2$2s$^2$2p$^2$-1s$^2$2s$^2$2p3d&12.259$^a$&12.247($^{+3}_{-2}$)&-290$^{+70}_{-50}$&4.2($^{+8}_{-9}$)&$<$5.4&14(3)&$<$14&10(3)$^d$&$\leq$10 \\
Fe XXI 1s$^2$2s$^2$2p$^2$-1s$^2$2s$^2$2p3d&12.285$^a$&12.304(2)&460$\pm$50&22(1)&$<$6.8&75(5)&$<$18&12(2)&$\leq$1.6 \\
Fe XXI 1s$^2$2s$^2$2p$^2$-1s$^2$2s$^2$2p3d&12.422$^c$&12.438(2)&390$\pm$50&3.9(8)&$<$2.9&14(3)&$<$8.1&3.0($^{+9}_{-8})$$^d$&$\leq$1.6 \\
Fe XX 1s$^2$2s$^2$2p$^3$-1s$^2$2s2p$^3$3p&12.576$^c$&12.583($^{+2}_{-3}$)&170$^{+50}_{-70}$&7(1)&$<$3.6&26(4)&$<$10&19($^{+6}_{-4}$)&$\leq$5.2 \\
Fe XX 1s$^2$2s$^2$2p$^3$-1s$^2$2s$^2$2p$^2$($^3$P)3d&12.82$^a$&12.844(2)&560$\pm$50&26(2)&$<$6.5&100(6)&$<$20&34$^{+16}_{-10}$&$\leq$1.0\\
Fe XX 1s$^2$2s$^2$2p$^3$-1s$^2$2s$^2$2p$_{1/2}$2p$_{3/2}$3d&12.912$^c$&12.914(3)&50$\pm$70&12(2)&$<$4.7&47(6)&$<$14&33$^{+11}_{-7}$&$\leq$5.9 \\
Fe XX 1s$^2$2s$^2$2p$^3$-1s$^2$2s$^2$2p$_{1/2}$2p$_{3/2}$3d&12.965$^c$&12.953($^{+3}_{-2}$)&-280$^{+70}_{-50}$&12(1)&$-$&48(6)&$-$&56$^{+25}_{-14}$&$-$ \\
Ne IX 1s$^2$-1s2p&13.448$^b$&13.448($^{+6}_{-5}$)&0$^{+130}_{-110}$&9(2)&$<$8.5&38(9)&$<$30&6($^{+3}_{-2}$)&$\leq$4.1 \\
Fe XIX 1s$^2$2s$^2$2p$^4$-1s$^2$2s$^2$2p$_{1/2}$2p$^2_{3/2}$3d&13.479$^c$&13.482(3)&70$\pm$70&5(1)&$<$3.3&20($^{+6}_{-5}$)&$<$12&1.0($^{+4}_{-3}$)$^d$&$\leq$0.5 \\
Fe XIX 1s$^2$2s$^2$2p$^4$-1s$^2$2s$^2$2p$^3$($^2$D)3d&13.518$^c$&13.523(3)&110$\pm$70&16(2)&$-$&70($^{+8}_{-9}$)&$-$&24$^{+16}_{-9}$&$-$ \\
Fe XVIII 1s$^2$2s$^2$2p$^5$-1s$^2$2s$^2$2p$^4$($^1$D)3d&14.203$^a$&14.220(3)&360$\pm$40&6(1)&$<$4.0&32(7)&$<$18&4($^{+3}_{-2}$)$^d$&$\leq$1.5\\
Fe XIX 1s$^2$2s$^2$2p$^4$-1s$^2$2s$^2$2p$^3$($^2$P)3s&14.60$^a$&\raisebox{-1.5ex}[0cm][0cm]{14.608(5)$^e$}&\raisebox{-1.5ex}[0cm][0cm]{100$\pm$100}&\raisebox{-1.5ex}[0cm][0cm]{19(3)}&\raisebox{-1.5ex}[0cm][0cm]{20(4)}&\raisebox{-1.5ex}[0cm][0cm]{81$\pm$13}&\raisebox{-1.5ex}[0cm][0cm]{73$\pm$14}&\raisebox{-1.5ex}[0cm][0cm]{200$^{+160}_{-90}$}&\raisebox{-1.5ex}[0cm][0cm]{170$^{+120}_{-60}$}\\
Fe XVIII 1s$^2$2s$^2$2p$^5$-1s$^2$2s$^2$2p$^4$($^3$P)3d&14.610$^a$&&&&&&&\\
\hline
\end{tabular}
Notes. --- Results for Periods I and II are both shown for comparison. The
errors in parentheses indicate uncertainty in the last digit of the
measurement; 1$\sigma$ errors are shown. Negative flux or EW upper limits
(indicating emission) are not shown. \\
a Verner et al. (1996); b Behar et al. (2002); c ATOMDB 1.3.3; d Unresolved; 
e The two transitions are equally probable. The average wavelength was used
to derive the Doppler-shift of the line.
\end{table}

\newpage
\begin{figure}
\psfig{figure=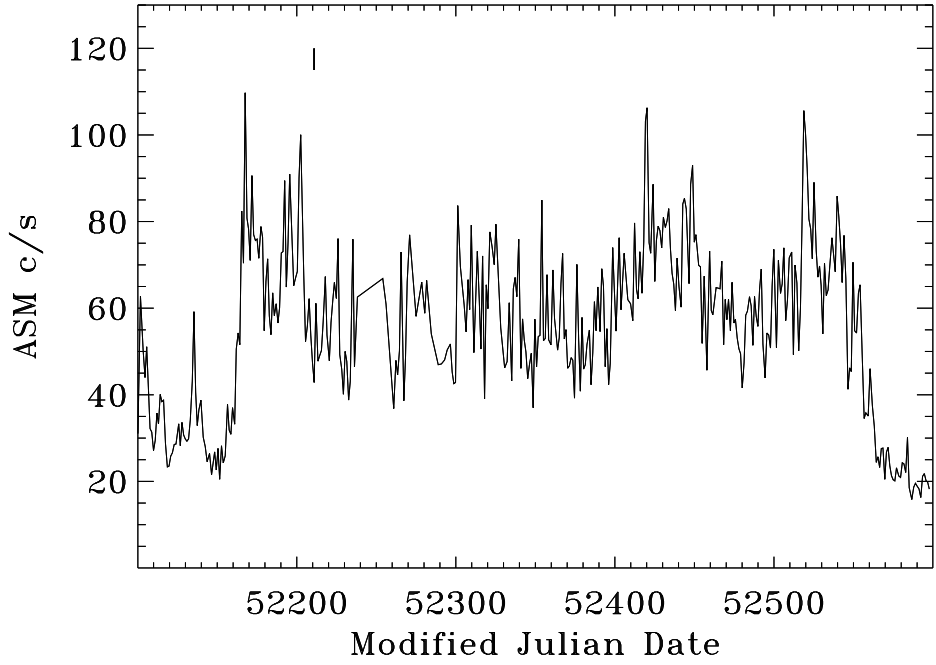,width=5.5in} 
\caption{Daily-averaged ASM Light Curve of Cyg X-1 during the 2001 state
transition. The vertical line indicates the time of the {\em Chandra} 
observation. }
\end{figure}

\begin{figure}
\psfig{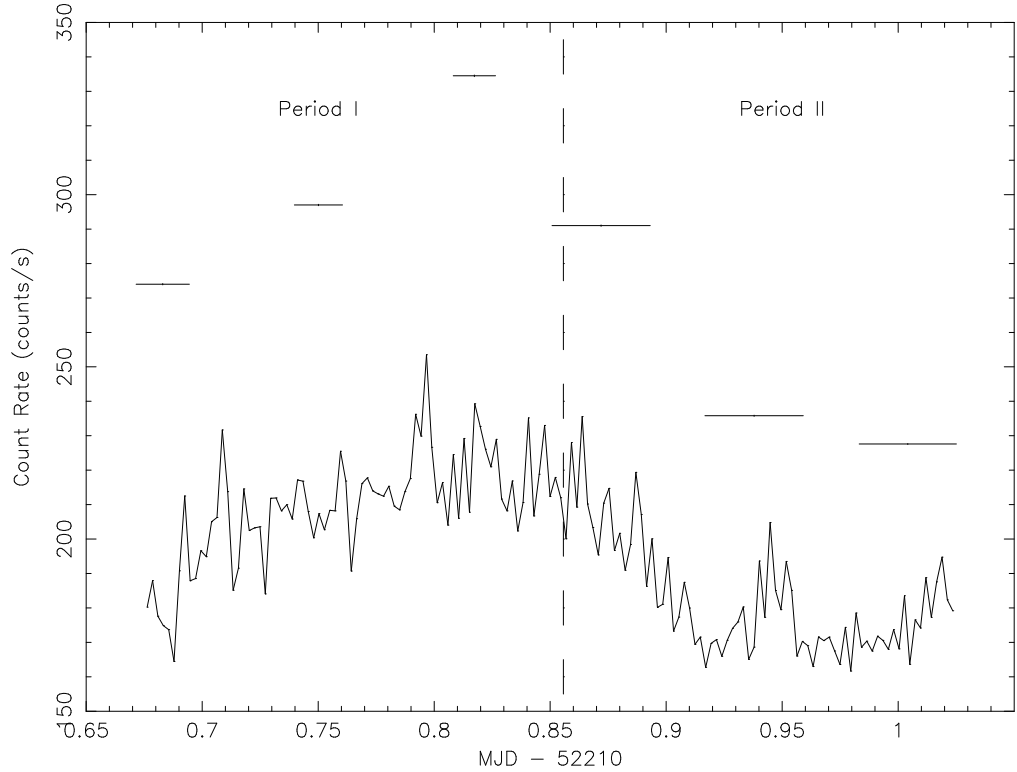} 
\caption{X-ray Light Curves of Cyg X-1. The solid curve shows data from the 
MEG first order, while the horizontal bars show the average count rates from 
PCU \#2. The error bars are negligible in both cases. The dashed line defines 
the two time periods for subsequent analyses (see text). }
\end{figure}

\begin{figure}
\psfig{figure=f3.eps,width=5.5in} 
\caption{MEG first-order spectrum of Cyg X-1 for Period I. No binning was
applied. The presence of absorption lines are apparent. The identifications 
of the lines are shown. Note that the emission-like features at 6.74 {\AA} 
and 7.96 {\AA} are likely instrumental (see text). The solid line shows
the best-fit to the (local) continuum. }
\end{figure}

\begin{figure}
\psfig{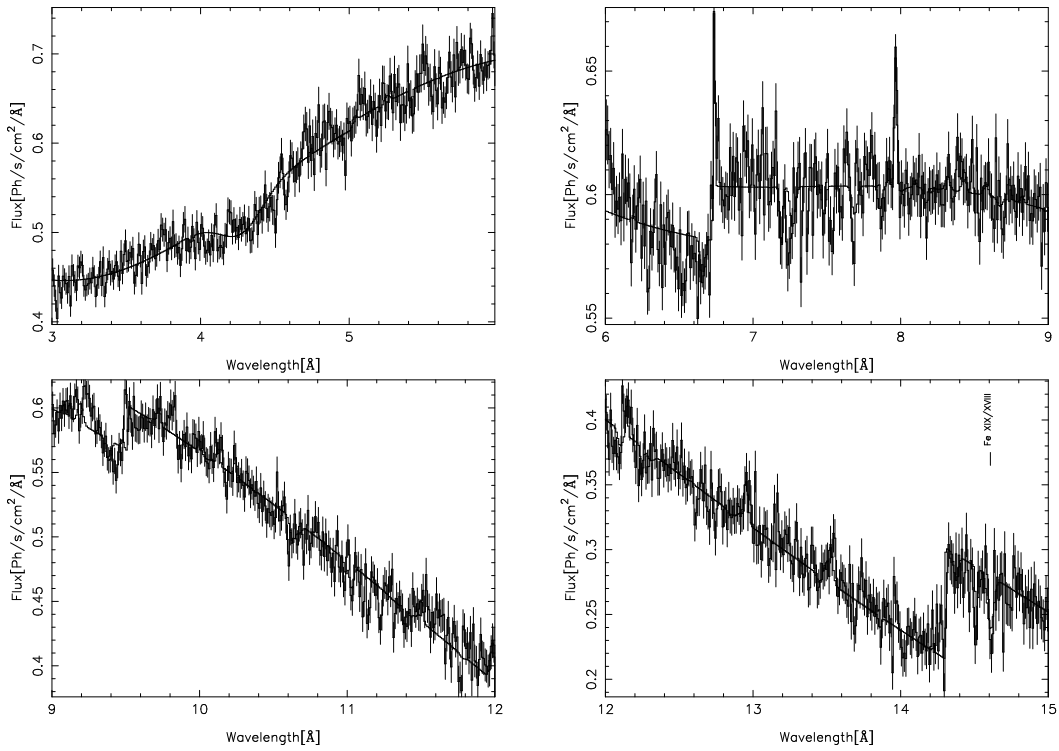} 
\caption{As in Fig.~3, but for for Period II. Note the absence of nearly all
the absorption lines seen in Fig.~3. }
\end{figure}

\begin{figure}
\psfig{figure=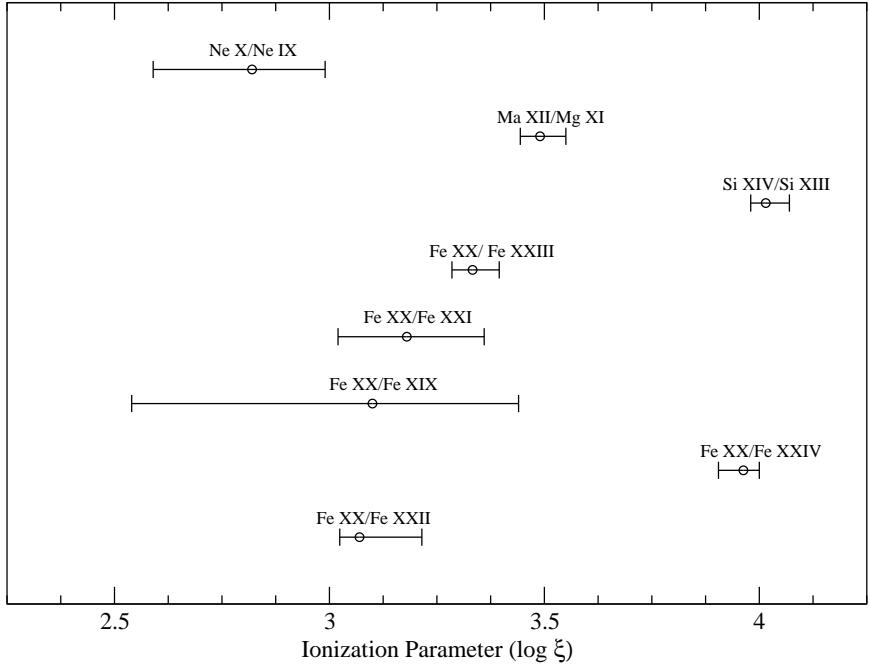,width=4.5in,angle=-90} 
\caption{Allowed ranges of the ionization parameter, each of which is inferred
from the ratio of the average densities of two ions of the same element. }
\vspace{0.2in}
\end{figure}

\begin{figure}
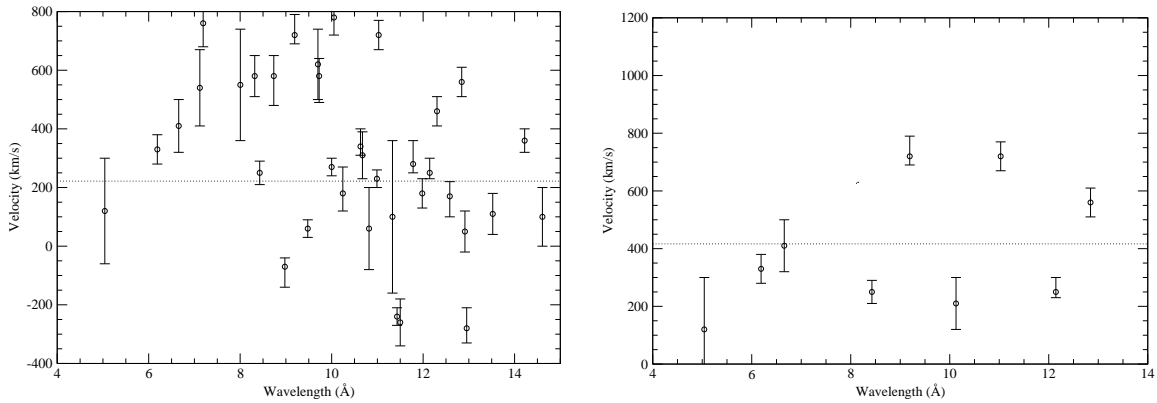

\psfig{figure=f6_left.eps,width=2.9in} \hspace{0.1in}
\psfig{figure=f6_right.eps,width=2.9in}
\caption{Inferred Doppler shift of the selected absorption lines, ({\it left})
all the lines with a significance above $5\sigma$ and ({\it right}) only the 
lines that were used by Marshall et al. (2001). The dotted line shows the 
average Doppler velocity in both cases. } 
\end{figure}

\begin{figure}
\psfig{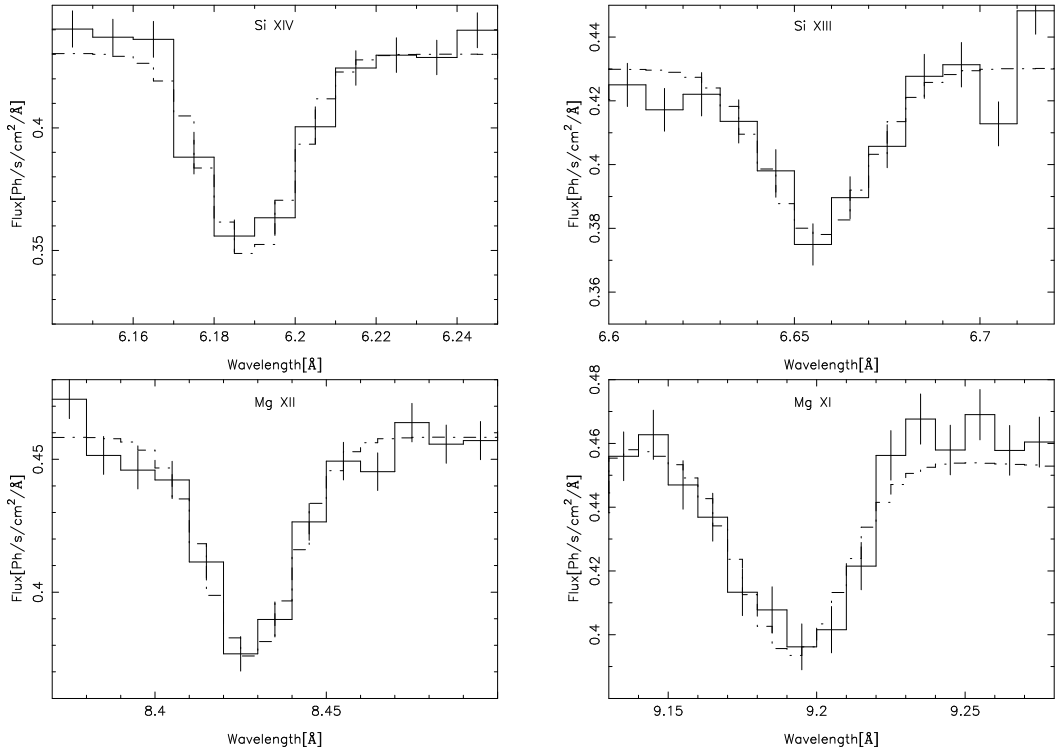} 
\caption{Profiles of the selected absorption lines. In each case, the 
dot-dashed histogram shows a fit to the profile with a Gaussian function. }
\end{figure}

\end{document}